\documentclass[conference]{IEEEtran}
\IEEEoverridecommandlockouts

\usepackage{cite}
\usepackage{amsmath,amssymb,amsfonts}
\usepackage{algorithmic}
\usepackage{graphicx}
\usepackage{textcomp}
\usepackage{xcolor}
\usepackage{mathtools}
\def\BibTeX{{\rm B\kern-.05em{\sc i\kern-.025em b}\kern-.08em
    T\kern-.1667em\lower.7ex\hbox{E}\kern-.125emX}}

\usepackage{siunitx}
\usepackage{subfigure}
\usepackage{svg}
\usepackage{booktabs}

\DeclareSIUnit\year{y}
\DeclareSIUnit\eur{EUR}

\begin{document}

\title{On the Complementarity of Shared Electric Mobility and Renewable Energy Communities\thanks{This work was supported by Klimaan cooperative, the LECaaS, REINVENT and  EFES projects.}}

\author{
\IEEEauthorblockN{Noé Diffels$^{1,2}$, Julien Allard$^{1}$, Bertrand Cornélusse$^{2}$ and Zacharie De Grève$^{1}$\\}
\IEEEauthorblockA{$^{1}$Power Systems and Markets Research Group, University of Mons, Belgium \\ $^{2}$Montefiore Institute, University of Liège, Belgium} 
}

\maketitle

\begin{abstract}
Driven by the ongoing energy transition, shared mobility providers are emerging actors in power systems aiming to shift combustion-based vehicles towards electric ones. Meanwhile Energy Communities are deployed to promote investment in distributed renewable production and enhance the local usage of it. The complementarity in their electrical demand, enhanced by a coordinated operational planning, can help both actors reduce the electricity supply cost. Considering this original collaboration, this paper presents a Mixed-Integer Quadratic Programming problem which jointly optimizes the EC members and EVs flexibility usage to take advantage of the local production. Besides economic benefits comparison, authors analyses the impact of grid tariffs and bi-directional charging on the  distribution network. Results from a Belgian mobility case study show that coordination can help reducing the yearly cost up to \textbf{15.6\%} compared to their stand-alone situation and that it may reduce by \textbf{30.8\%} the stress on the substation transformer when subject to peak penalties from the grid operator.
\end{abstract}


\section{Introduction}
\noindent Driven by the European climate policies and the technological advances that have decreased the cost of batteries \cite{IEA_EV}, the Electric Vehicles (EV) market is experiencing a sustained growth in recent years. However, this rapid shift in the mobility paradigm raises challenges for the current power systems \cite{deb2017review}. 
To mitigate these, many smart charging strategies of private EVs have been proposed (e.g., charging at the office to balance solar production \cite{mouli2016system}). 
In addition, besides private EVs, shared mobility has emerged as a key strategy to disconnect vehicle investment and usage while optimizing vehicle operation \cite{machado2018overview}. From a grid usage point of view, the erection of shared EV stations in high-density residential area may mitigate the intermittency of local renewable production. Indeed, despite the increased uncertainty in shared mobility demand compared to private cars, charging intervals are usually uniformly spread over the day. One of the primary objectives of Shared Mobility Providers (SMPs) is to offer to their end-users the possibility of travelling at low cost. Hence, they are interested to access cheap electricity to supply the EV fleet.

In the meantime, concept of local electricity sharing between end-users has been conceptualized by the EU through the Clean Energy Package \cite{EU} as the so called Renewable Energy Communities (REC). This new market paradigm enables end-users to gather and exchange renewable energy locally in any form of energy carrier. Doing so, they bypass the traditional retailer to supply part of their energy demand at more advantageous prices \cite{zanvettor2022optimal}. This mutualisation of resources aims to provide economic, environmental and social advantages to participants \cite{ref1a}. This new framework fosters SMPs integration, which could take advantage of the local production surplus to charge its vehicles. Additionally, by coordinating with the local end-users, SMPs may improve the technical and economic efficiency of its charging strategies, and even provide flexibility services to system operators \cite{da2024electric, allard2024quantifying}. 



\subsection{Brief review of the related work}

\noindent Research investigating the complementarity of mobility and/or shared mobility with REC operations is still relatively sparse. In \cite{zanvettor2022optimal}, a day-ahead planning problem for the management (i.e., EV-ride assignment) of a shared EV fleet within a REC is formulated. The bi-level problem is solved using a heuristic approach highlighting the reduced charging costs for fleet operators on small test cases (3 to 5 REC members). In \cite{velkovski2024framework}, authors extend the use of community-owned shared power stations to external users to increase the consumption of the REC local production. Hence, they demonstrate the benefits brought to the members when external users charging are allowed. In addition, several recent contributions explicitly quantify and exploit the flexibility provided by EVs within REC frameworks \cite{da2024electric, allard2024quantifying}. They show how EV charging can be controlled to increase the community social and economic benefits. Overall, current literature mainly focuses on the economic benefits either of EC members or EV fleet charging, disregarding the physical impacts on the local distribution network. Besides, the repartition of the benefits between the EC and SMPs has not been addressed yet while previous works showed that adding a new member in an existing community may have a negative impact on current members \cite{allard2025dyn}.

\subsection{Contributions and paper organization}
\noindent This paper proposes an optimisation framework for the operational coordination of a shared EV fleet with the production and flexible resources of a REC. The developed model quantifies the economic and grid performances of their joint operation. The main paper contributions are:
\begin{enumerate}
    \item An original Mixed-Integer Quadratic Programming (MIQP) problem which jointly defines the optimal EC flexibility schedule, and SMP charging and ride-assignment strategies to minimise the electricity cost;
    \item The quantification of the technical impacts on the feeder substation when peak-power penalties are applied by the DSO or bi-directional charging is considered.
\end{enumerate}
We further study the impact of energy distribution rules on the respective savings of ECs and SMPs. The mobility demand used for the case study is extracted from a real historical database of Klimaan, a Belgian SMP \cite{REINVENT}.
\section{Optimization Framework and Key Assumptions}
\label{sec:Models}
\subsection{Energy Community Model}
\label{sec:EC}
\noindent This study considers a domestic REC equipped with a joint PV installation. The local production is made available to its members who may activate flexibility in their electricity demand. In the stand-alone configuration, the REC aims to minimize the collective bill by optimizing the flexible loads to take advantage of local generation. The REC problem can be expressed, in general terms, as:
\begin{align} 
     \min_{\Omega_{EC}} \quad & \text{$C^{EC}$} \label{SMP-CERmodel}\\
     \textrm{s.t.}\quad
     & \text{Power balance constraint (\ref{eq:Bal1})}\nonumber\\
     & \text{Flexibility and PV constraints (\ref{eq:PV})-(\ref{eq:sumflex})}\nonumber
\end{align}
\noindent The set of decision variables of the EC is $\Omega_{EC} = \{e^{ret}_{t}, i^{ret}_{t},p^{PV}_{t},p^{flex}_{b,t}\}$, which are respectively the electricity exported to and imported from the retailer, the controlled PV output and the members load demand. The objective of the EC, comprising $b\in\mathcal{B}$ members over $t\in\mathcal{T}$ time steps of duration $\Delta t$, is to minimize the total cost $C^{EC}$, defined as:
\begin{align}
    &C^{EC} &= \underbrace{\sum_{t\in\mathcal{T}} \lambda^{ret,imp}_{t}i^{ret}_{t}\Delta t}_{C^{ret,EC}} - \underbrace{\sum_{t\in\mathcal{T}} \lambda^{ret,exp}_{t}e^{ret}_{t}\Delta t}_{R^{ret}}\nonumber\\ & & + \underbrace{\sum_{b\in\mathcal{B},t\in\mathcal{T}} \alpha \ (p^{flex}_{b,t} - P^{load}_{b,t})^2}_{C^{disc}} \label{Cost_EC}
\end{align}
Which is composed of the retailer bill (i.e., costs $C^{ret,EC}$, and revenues $R^{ret}$) for the volumes traded at prices $\lambda^{ret,imp}_t$ and $\lambda^{ret,exp}_t$, for imports and exports respectively. The retailer is considered unique for all members. In addition, a quadratic discomfort cost $C^{disc}$ is associated to the deviation of end-users consumption from a baseline load profile $P^{load}_{b,t}$. A discomfort reluctance factor $\alpha$ reflects the inconvenience experienced by members when altering their consumption patterns. The EC operation is subject to the following set of constraints:
\begin{align}
    i^{ret}_{t} + p^{PV}_{t} &= e^{ret}_{t} + \sum_{b\in\mathcal{B}}p^{flex}_{b,t} &\forall t \in \mathcal{T} \label{eq:Bal1} \\
    p^{PV}_{t} &\leq \bar{P}^{PV}_{t} &\forall t \in \mathcal{T} \label{eq:PV} \\
    p^{flex}_{b,t} &\leq \bar{P}^{load}_{b,t} &\forall b \in \mathcal{B}, \forall t \in \mathcal{T} \label{eq:flexlim1}\\
    p^{flex}_{b,t} &\geq \underline{P}^{load}_{b,t} &\forall b \in \mathcal{B}, \forall t \in \mathcal{T} \label{eq:flexlim2}\\
    \sum_{t\in\mathcal{T}_d}p^{flex}_{b,t} &= \sum_{t\in\mathcal{T}_d}P^{load}_{b,t} &\forall b \in \mathcal{B}, \forall d \in \mathcal{D} \label{eq:sumflex}
\end{align}
\noindent The power balance at the EC level is expressed in (\ref{eq:Bal1}), where the controlled joint PV production $p^{PV}_{t}$ is capped by the solar potential $\bar{P}^{PV}_{t}$ in (\ref{eq:PV}). Equations (\ref{eq:flexlim1}) and (\ref{eq:flexlim2}) restrict the power deviation around the baseline, while (\ref{eq:sumflex}) enforces the daily consumption of each member to be fixed, where $\mathcal{T}_d$ represents the set of time periods within day $d\in\mathcal{D}$.

\subsection{Shared Mobility Provider Model}
\label{sec:SMP}
\noindent The SMP aims to satisfy a deterministic mobility demand estimated for its operation area by assigning EVs from its fleet to ride requests. To do so, the SMP objective is to minimize the electricity cost for charging the EV batteries in order to offer affordable price for its service. The flexibility lever when managing an EV fleet lies in the charging pattern of the vehicles when connected to the deposit charging stations. In addition, the Vehicle-to-Grid (V2G) feature or a battery storage system (BSS) coupled with the charging station can be included to enhance the flexibility potential. In this framework, the bidirectional exchanged power at the charging station can be controlled to perform an arbitrage on the retailer dynamic prices. The SMP problem can be summarized as follows:
\begin{align}
     \min_{\Omega_{SMP}} \quad & C^{SMP}\\
     \textrm{s.t.}\quad
     & \text{Power balance constraint (\ref{eq:Bal2})}\nonumber\\
     & \text{Ride-EV assignment constraints (\ref{eq:ride-EV1})-(\ref{eq:ride-EV3})}\nonumber\\
     & \text{EV and CS operational constraints (\ref{eq:Op1})-(\ref{eq:Op10})}\nonumber     
\end{align}

\noindent The set of decision variables of the SMP is $\Omega_{SMP} = \{e^{ret}_{t}, i^{ret}_{t}, \delta^{use}_{n,r}, \delta^{state}_{n,t}, p^{CS}_{s,t}, p^{EV}_{n,t}, s^{EV}_{n,t}, e^{away}_{n,r}, p^{BSS}_{t}, s^{BSS}_t\}$, with, respectively, the retailer exports and imports, the EV-ride assignment decision, the EV state (i.e., at the charging station or on the road), the charging station power exchange, the EV power exchange, the EV state-of-charge (SOC), the energy charged during a ride, the static BSS power exchange and SOC. The overall bill paid by the shared mobility operator, $C^{SMP}$, is defined as:
\begin{align}
    &C^{SMP} &=\nonumber \underbrace{\sum_{t\in\mathcal{T}} \lambda^{ret,imp}_{t}i^{ret}_t\Delta t}_{C^{ret,SMP}} + \underbrace{\sum_{\mathclap{\substack{\quad n\in\mathcal{N}^{EV}\\r\in\mathcal{R}}}} \lambda^{away}e^{away}_{n,r}}_{C^{away}} \nonumber \\
    & &+ \underbrace{\pi^{uns}\sum_{r\in\mathcal{R}} e^{uns}_r}_{C^{uns}}
\end{align}
Both first terms represent grid and away charging costs. The latter is priced at $\lambda^{away}$ and may happen if a trip is too long or the charging period is too short. Additionally, an opportunity cost, $C^{uns}$, is considered to represent the revenues missed for not serving certain ride demand $d_r$, $r\in\mathcal{R}$, with $e^{uns}_r = d_r(1-\sum_{n\in\mathcal{N}^{EV}} \delta^{use}_{n,r})$, valued at price $\pi^{uns}$.

\subsubsection{Power balance constraint}
The power balance equation, considering only grid-to-vehicle, is simply given by: 
\begin{align}
    i^{ret}_{t} =  p^{CS}_{t}\label{eq:Bal2}
\end{align}

\subsubsection{Ride-EV assignment constraints}
The proposed shared EV fleet planning problem embeds the following sets of time-overlapping rides-EV assignment constraints: 
\allowdisplaybreaks{
\begin{align}   
    &\hspace{-0.4cm}\sum_{\quad n\in\mathcal{N}^{EV}} \hspace{-0.4cm}\delta^{use}_{n,r} \leq 1, 
    & \forall r \in \mathcal{R}, \label{eq:ride-EV1}\\
    &\delta^{use}_{n,r_1} + \delta^{use}_{n,r_2} \leq 1 
     &\forall n\in\mathcal{N}^{EV}, \forall(r_1,r_2)\in\mathcal{R}^{2'}, \label{eq:ride-EV2}\\
   &\delta^{state}_{n,t} = \sum_{r\in\mathcal{R}_t} \delta^{use}_{n,r}, 
    & \forall n\in\mathcal{N}^{EV},\forall t \in \mathcal{T}. \label{eq:ride-EV3}
\end{align}
}

\noindent First, \eqref{eq:ride-EV1} ensures that only one car can be assigned (i.e., $\delta^{use} = 1$) to each ride. Then, \eqref{eq:ride-EV2} checks that the same car is not assigned to overlapping rides defined by the set $\mathcal{R}^{2'}:= \{(r_1,r_2) \in \mathcal{R} \times \mathcal{R} | t^{{dep}}_{r_1}\leq t^{{ret}}_{r_2}, \; t^{{dep}}_{r_2}\leq t^{{ret}}_{r_1}\}$. Finally, \eqref{eq:ride-EV3} sets the state of EVs to 1 if they are on a ride and 0 if they are at the charging station, with $\mathcal{R}_t:= \{r \in \mathcal{R} | t^{{dep}}_{r}\leq t \leq t^{{ret}}_{r}\}$.

\subsubsection{EV and CS operational constraints}
Electric vehicles are modelled as bidirectional batteries with intermittent availability whose exchanged power can be adjusted when standing at the deposit stations before the next ride:
\allowdisplaybreaks{\begin{align}  
     &0 \leq p^{{EV,ch/dch}}_{n,t} \leq \overline{P}^{EV}_n(1 -\delta^{state}_{n,t}),  \hspace{-4cm}\nonumber\\
     &&\forall n\in\mathcal{N}^{EV},\forall t\in\mathcal{T}, \label{eq:Op1}\\
    &s^{EV}_{n,t} = s^{EV}_{n,t-1} 
    + \big(\eta^{EV} p^{{EV,ch}}_{n,t} - p^{{EV,dch}}_{n,t}\big)   \Delta t  \hspace{-6cm}
    \nonumber\\
    & \qquad - \hspace{-0.4cm}\sum_{\quad r \in \mathcal{R}^{ret}_t} \hspace{-0.4cm} d_r \,\delta^{use}_{n,r} +  \hspace{-0.4cm}\sum_{\quad r \in \mathcal{R}^{ret}_t} \hspace{-0.4cm}e^{away}_{n,r},\hspace{-0.1cm}& \forall n\in\mathcal{N}^{EV},\forall t\in\mathcal{T},\label{eq:Op2}\\
    &\underline{E}^{EV}_n \hspace{-0.2cm}\leq s^{EV}_{n,t} \leq \overline{E}^{EV}_n, 
    & \forall n\in\mathcal{N}^{EV},\forall t\in\mathcal{T},\label{eq:Op3}\\
    &s^{EV}_{n,t^{dep}_r} \geq \alpha^{dep}\overline{E}_n,
    & \forall n\in\mathcal{N}^{EV},\forall r\in\mathcal{R},\label{eq:Op4}\\
    &e^{away}_{n,r} \leq \overline{E}^{away}_r \delta^{use}_{n,r}, & \forall n\in\mathcal{N}^{EV},\forall r\in\mathcal{R}.\label{eq:Op5}\\
    &0\leq p^{{BSS,ch/dch}}_{t} \leq \overline{P}^{BSS},  &\forall t\in\mathcal{T}, \label{eq:Op6}\\
    &s^{BSS}_{t} = s^{BSS}_{t-1} +\hspace{-0.1cm} \big(\eta^{BSS} p^{BSS,ch}_{t}\hspace{-0.3cm}-p^{BSS,dch}_{t}\big) \Delta t, \hspace{-2cm}&\forall t\in\mathcal{T},\label{eq:Op7}\\
    &\underline{E}^{BSS} \hspace{-0.1cm}\leq s^{BSS}_{t} \leq \overline{E}^{BSS}, & \forall t\in\mathcal{T},\label{eq:Op8}\\
    & p^{CS}_{t} = p^{BSS}_t +\hspace{-0.5cm} \sum_{\quad n\in\mathcal{N}^{EV}} \hspace{-0.4cm}p^{EV}_{n,t},  & \forall t\in\mathcal{T},\label{eq:Op9}\\
    &-\overline{P}^{CS} \leq p^{CS}_{t} \leq \overline{P}^{CS}, & \forall t\in\mathcal{T}. \label{eq:Op10}
\end{align}
}

\noindent The power fed to, $p^{{EV,ch}}_{n,t}$, or extracted from EVs, $p^{{EV,dch}}_{n,t}$, at deposit charging station are limited by their availability and their power ratings, $\overline{P}^{EV}_{n}$ via \eqref{eq:Op1}. The EV SOC evolves with the power exchanged at the CS considering an efficiency $\eta$, but also the consumed energy during its usage by customers, and the energy filled during the trip in \eqref{eq:Op2}. Both terms are included in the evolution equation at the return time (i.e., $\mathcal{R}^{ret}_t:= \{r \in \mathcal{R} |t = t^{{ret}}_{r}\}$) as the rides temporal dynamics are not modelled. The EV's SOC is bounded by its capacity, $\overline{E}_n$, and a minimum level, $\underline{E}_n$, using \eqref{eq:Op3}. \eqref{eq:Op4} imposes that, at departure time, $t^{dep}_r$, a minimum fraction of the battery, $\alpha^{dep}$, must be available. Finally, the energy charged away is limited in \eqref{eq:Op5} based on the typical public charging station rating, $\overline{P}^{away}$ and the trip duration as $\overline{E}^{away}_r=\overline{P}^{away} \cdot (t^{\text{ret}}_r - t^{\text{dep}}_r)\Delta t$. Similarly, constraints \eqref{eq:Op6} to \eqref{eq:Op8} represent the operation of a static BSS. Finally, constraint \eqref{eq:Op9} computes the net power that is exchanged with the distribution network at the deposit location, which is bounded by the CS power ratings in \eqref{eq:Op10}.

\subsection{Coordinated Model}
\label{sec:coord_model}
\noindent The SMP and EC operations can be jointly organized to take advantage of each other resources and generate benefits for both actors. This coordinated model combines the stand-alone formulations and is defined as follows:
\begin{align}
     \min_{\Omega} \quad & C^{tot}\\
     \textrm{s.t.}\quad
    & \text{Power balance constraint (\ref{eq:Bal3})}\nonumber\\
     & \text{Flexibility and PV constraints (\ref{eq:PV})-(\ref{eq:sumflex})}\nonumber\\
     & \text{Ride-EV assignment constraints (\ref{eq:ride-EV1})-(\ref{eq:ride-EV3})}\nonumber\\
     & \text{EV and CS operational constraints (\ref{eq:Op1})-(\ref{eq:Op10})}\nonumber
\end{align}

\noindent $\Omega= \Omega_{EC} \cup \Omega_{SMP}$, is the full set of decision variables. The total cost of electricity can be expressed as:
\begin{align}
    &C^{tot} = C^{ret,tot} -R^{ret} + C^{disc} +C^{away}+C^{uns}\label{eq:costs_coord}
\end{align}
Finally, (\ref{eq:Bal1})~and~(\ref{eq:Bal2}) are merged into (\ref{eq:Bal3}), to consider the charging of the shared vehicles within the community balance.
\begin{align}
    i^{ret}_{t} + p^{PV}_{t} = e^{ret}_{t} + \sum_{b\in\mathcal{B}}p^{flex}_{b,t} + p^{CS}_{t} &&\forall t \in \mathcal{T} \label{eq:Bal3} 
\end{align}
\section{Energy allocation and individual bills}
\label{sec:kor}
\noindent After the coordinated model is solved and the optimal flexibility usage is obtained, their respective savings will depend on the allocation of the local PV production between the two entities. The individual bills of the SMP and the EC can be computed as follows:
\begin{align}
    &C^{SMP, coord} = C^{SMP} + \sum\limits_{t\in\mathcal{T}} \lambda^{EC}_t i^{com,SMP}_t \Delta t \hspace{-0.3cm}\\
    &C^{EC, coord} = C^{EC} - \sum\limits_{t\in\mathcal{T}} \lambda^{EC}_t i^{com,SMP}_t \Delta t\\
    &p^{CS}_t = i^{ret,SMP}_t+i^{com,SMP}_t & \forall t \in \mathcal{T}
\end{align}
\noindent With $i^{ret,SMP}_t$ and $i^{com,SMP}_t$, the energy that is traded by the SMP with its retailer and with the EC at $\lambda^{EC}_t$, respectively.
In this work, three repartition mechanisms are compared in the case of unidirectional flow at the charging station:
\begin{enumerate}
    \item EC prioritization over SMP:\\ $i^{com,SMP}_t = \min (p_t^{CS},\min(0,p^{PV}_{t}- \sum_{b\in\mathcal{B}}p^{flex}_{b,t}))$
    \item Prorate consumption sharing:\\ $i^{com,SMP}_t = \min \Big(p_t^{CS},p^{PV}_{t}\cdot\frac{p_t^{CS}}{ p^{CS}_t+\sum\limits_{b\in\mathcal{B}}p^{flex}_{b,t}}\Big)$
    \item Hybrid sharing:\\
    $i^{com,SMP}_t = \min (p_t^{CS},\frac{p^{PV}_t}{2}+\min(0,\frac{p^{PV}_{t}}{2}- \sum\limits_{b\in\mathcal{B}}p^{flex}_{b,t}))$
\end{enumerate}
 
\section{Historical Mobility Data Extraction}
\label{sec:mobility}

\noindent This study relies on a real historical dataset for the mobility demand, provided by a Belgian SMP \cite{REINVENT}. This dataset contains around 7600 recorded trips and 4500 charging events from a fleet of 35 shared EVs. Building on this dataset, scenarios of specific mobility demand of shared vehicles are obtained for the case study further presented in Section~\ref{sec:results}. 
\begin{figure}
    \centering
    \includegraphics[width=\linewidth]{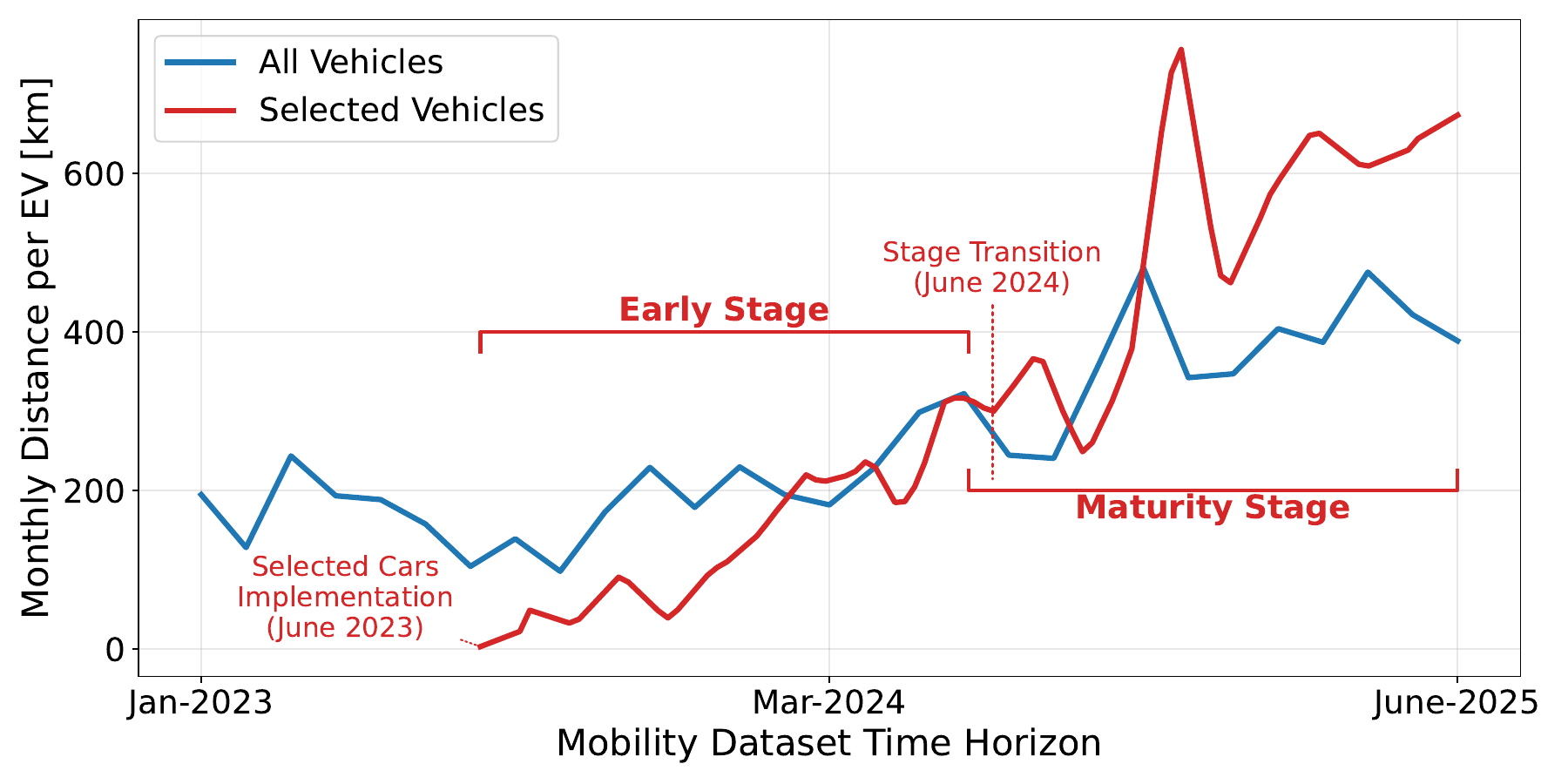}
    \caption{Monthly distance travelled by the shared vehicles.}
    \label{fig:EV_trend}
\end{figure}

\noindent A key aspect of shared mobility, compared to private cars, is the long-term dynamic of vehicles adoption. As shown in Figure~\ref{fig:EV_trend}, two main stages can be identified after the implementation of a new asset: the early stage and the maturity stage. In the early stage, the EV usage is often sub-optimal as adoption still needs to grow. Once the SMP has gained sufficient visibility of the new mobility service, or that there is a shift in inhabitants' habits, it enters the maturity phase. In this study, two cars covering a two-year period (from July 2023 to July 2025) are extracted from the whole fleet. The first year represents the early stage and the second year the maturity stage. Figure~\ref{fig:EV_trend} shows that the long-term adoption trend of those cars (in red) follows the general dynamic (in blue) with a lower travelled distance during the early stage and a slightly higher level of maturity usage. In terms of technical properties, the extracted EV are a \textit{Renault ZOE} and an \textit{Opel CORSA}, with battery capacities of 50kWh. In addition, one deposit charging station is considered in this study with a maximum power charge (and discharge) of 11kW.


\begin{figure}
    \centering
    \includegraphics[width=\linewidth]{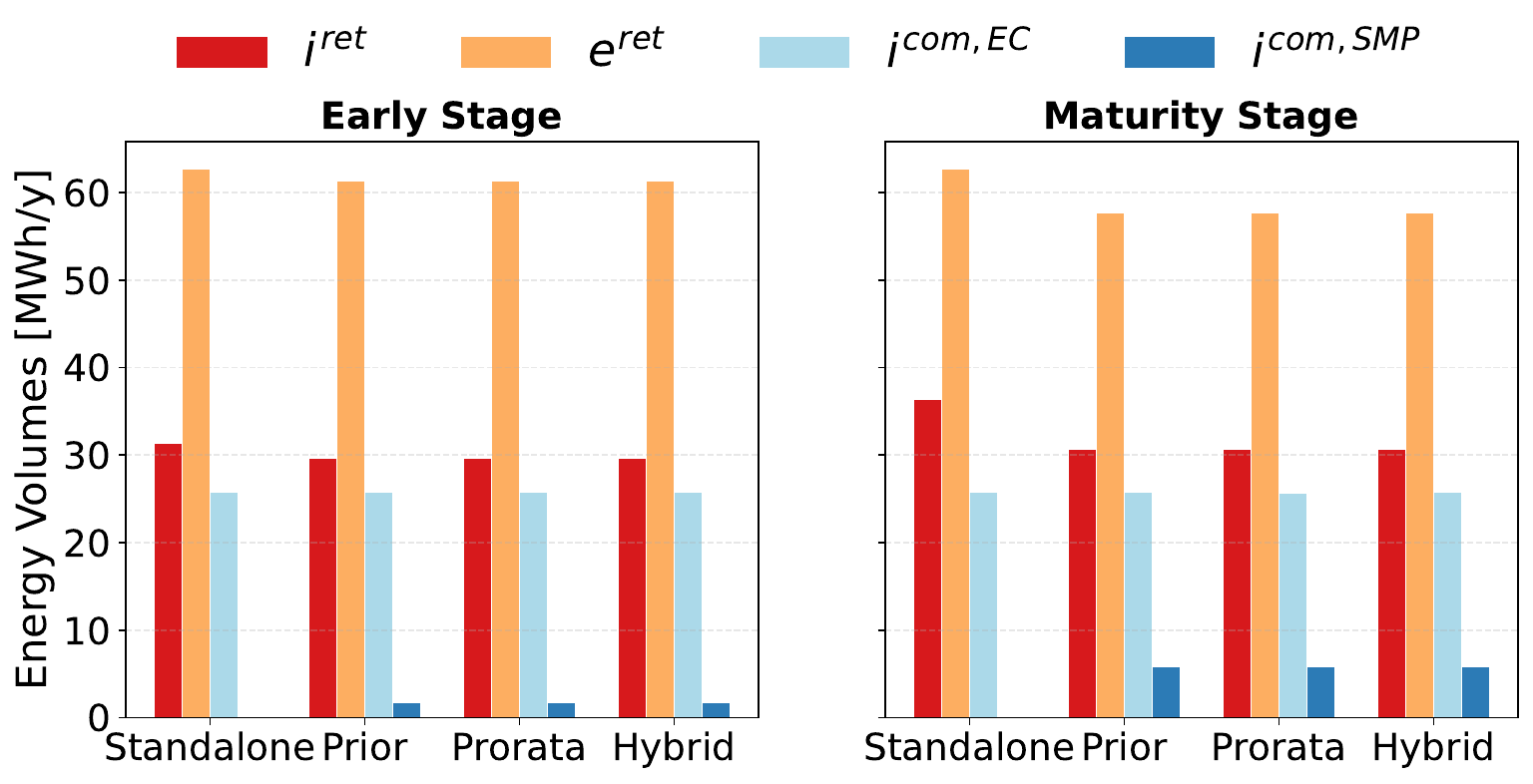}
    \caption{Energy volumes repartition for different sharing mechanisms for the two periods of EV integration.}
    \label{fig:results_kor}
\end{figure}

\section{Case Study and Results}
\label{sec:results}
\noindent The REC considered in this work is composed of 20 households, among which 10 end-users are equipped with a total of 97 \SI{}{\kilo\watt} of PV installations. The household consumption profiles are flexible and generated using a Residential Stochastic Flexible load profiles generator \textit{ResFlex} \cite{resflex} with a yearly consumption of \SI{55}{\mega\watt\hour} for the whole community. Historical dynamic electricity import and export retail prices are derived from the belgian spot price \cite{BelPex} with an additional 0.099 €$/kWh$ for distribution and transportation grid usage \cite{VREG}. The electricity used to charge the EV when riding is priced at 0.45 €$/kWh$, which is the highest retailer import price with a $25\%$ margin. The mobility demand scenarios used are the early and maturity stages introduced in Section~\ref{sec:mobility}. If mobility requests are not satisfied, a virtual cost of $\pi_{uns} = 2$ €$/kWh$ is applied. While Section~\ref{sec:results_eco} focuses on the economic performance of the SMP-community coordination, Section~\ref{sec:results_grid} extends the analysis by quantifying technical impacts on the local network for different scenarios: with V2G capability or BSS, and with individual or collective DSO peak tariffs.

\subsection{Coordination creates economic benefits}
\label{sec:results_eco}
\noindent The REC-SMP coordination first yields to mutual economic benefits. However, these benefits will progressively increase as the shared EVs are adopted by end-users. Indeed, in the early phase, the coordinated model achieves a 7\% reduction in the total costs $C^{tot}$ compared to stand-alone configuration, decreasing the bill from 2,531€ to 2,354€. Then, when entering the maturity phase, the joint objective is 3,470€, representing a 15.66\% reduction compared to the combined stand-alone costs, 4,115 €. These results highlight that cost reductions are driven by the increase in local self-consumption under the coordinated scheme which is directly linked to the mobility demand volume. This limits the coordination advantage in early stage with low mobility demand. At the same time, the discomfort costs induced by shifted user load profiles decreases by 2\% when coordination is activated.

\noindent In this case, one challenge is to determine a fair allocation scheme to share the benefits among the community members and the SMP. Figure~\ref{fig:results_kor} shows that the three local PV production sharing mechanisms introduced in Section~\ref{sec:kor} lead to similar distribution of benefits among the two actors. This is because the SMP is a minor consumer relative to the community (EV charge represents 4.2\% and 14.8\% of the EC consumption during the early and maturity stage, respectively). Under all three mechanisms, the EC is able to capture a significant local PV production covering 46.7\% to 46.8\% of the EC load. By enabling the SMP to charge the EV with the remaining local PV production (i.e., EC prioritization), the exchanges with the retailer of both actors decrease and 879€ are redirected from the SMP to the EC over the whole period. The repartition of the coordinated model bill reduction brings slight reduction of the average electricity price to cover EC members consumption and the average mobility service fee compared to stand-alone model, from 0.035 to 0.029 €/kWh, and from 0.10 to 0.082 €/km. 

\subsection{Impact of Individual and Collective capacity tariffs}
\label{sec:results_grid}
\noindent This section evaluates the stress on the substation transformer considering that all 20 EC members households and the SMP are connected behind the same substation. The analysis considers first the results obtained in the previous section with only volumetric DSO cost. Subsequently, to reduce the degradation of the transformer at the substation, a DSO capacity tariff charged on the maximum power load, as in Flanders (BE) \cite{VREG} is considered. The cost charged by the DSO, $C^{peak}=\pi^{peak} p^{peak}$, is added to $C^{tot}$ in (\ref{eq:costs_coord}) with the peak power price $\pi^{peak}= 59$ €$/kW/year$.
In the EC framework, the DSO may either charge individual import peak power, $p^{peak} = \overline{p}^{CS} + \sum_{b \in \mathcal{B}} \overline{p}^{flex}_b$ with $\overline{p}^{flex}_b$ and $\overline{p}^{CS}$ the largest consumption level of end-users and SMP, or collective peak power, with $p^{peak} = \overline{i}^{ret}$, defined as the maximum import power measured at the transformer. Table~\ref{tab:grid_results} presents the  active power import peaks for the SMP, the maximum peak over EC members $\overline{p}^{flex}=\max(\overline{p}^{flex}_b)$, and the aggregate peak observed at the slack node for the three tariff structures.  

\begin{table}[htbp]
\centering
\caption{Grid performance indicators (kW) and Objective functions (€) by tariff structure and bi-directional models.}
\label{tab:grid_results}
\footnotesize
\setlength{\tabcolsep}{4pt}
\begin{tabular}{l rrrr rrrr}
\toprule

 & \multicolumn{4}{c}{\textit{Early Stage}} 
 & \multicolumn{4}{c}{\textit{Maturity Stage}} \\
 
\cmidrule(lr){2-5}\cmidrule(lr){6-9}
  & $\overline{p}^{CS}$ & $\overline{p}^{flex}$ & $\overline{i}^{ret}$ & Obj.
 & $\overline{p}^{CS}$ & $\overline{p}^{flex}$ & $\overline{i}^{ret}$ & Obj. \\
\midrule
Volumetric  & 11.00 & 9.11 & 24.11 & 2353     & 11.00 & 9.11 & 26.92 & 3470 \\
Individual  & 7.02  & 6.77 & 22.71 & 5775      & 11.00 & 6.77 & 26.74 & 7110 \\
Collective  & 11.00 & 9.11 & 18.62 & 3205     & 11.00 & 9.12 & 18.62 & 4347\\
\midrule
V2G coll.  & 11.00 & 9.05 & 15.07 & 499      & 11.00 & 9.15 & 17.58 & 2317\\
BSS coll.  & 11.00 & 9.03 & 20.94 & -383      & 11.00 & 9.04 & 25.01 & 889\\
\bottomrule
\end{tabular}
\end{table}

\noindent By leveraging the combined flexibility of the EVs and residential loads, the collective peak tariff reduces $\overline{i}^{ret}$ by 28.5\% and 30.8\% in early and maturity stages, respectively. The resulting $C^{peak}$ to achieve this grid relief are 826.74~€ and 837.20~€, representing 25.79\% and 19.25\% of the total objective. On the contrary, while offering lower individual peak but higher $\overline{i}^{ret}$, the individual tariff charges a substantially higher $C^{peak}$ of 3196.86~€ and 3361.32~€, reflecting the penalization of all 21 actors independently. Because the charge of the EV is low relative to the EC consumption, the individual tariff is the only mechanism able to reduce $\overline{p}^{CS}$ through a direct constraint at the EV charge and not on the EC-SMP load aggregation. However, it significantly increases user discomfort doubling the load-shifting cost relative to collective or volumetric cases and decreases local energy exchanges by 2\%.

\noindent To further increase the flexibility potential of the coordination scheme, V2G or BSS coupled with the charging station can be considered. 
With the collective capacity tariff, these models achieve the lowest total objectives across both stages, driven primarily by their ability to absorb peak demand through storage rather than only load shifting. The early phase exhibits significantly higher flexibility potential compared to the maturity phase. In the latter, higher mobility demand reduces the EV availability, thus shrinking the windows where the EV can act as a stationary battery. While the BSS offers a lower objective, the V2G allows to reduce the stress at the transformer. In terms of discomfort, the V2G and especially the BSS help to reduce the user discomfort by 46.4\% and 70.8\% compared to unidirectional charging collective tariff, while keeping the same level of served trips and charge at CS of the EV.

\section{Conclusions}

\noindent This paper demonstrates the economic and technical synergies arising from the coordinated operation of an EC and a SMP. Key findings show that these benefits depend on the EV integration stage: while early stage deployment enables flexible grid support through longer station-connected periods, maturity stage generates greater mutual economic benefit through higher local consumption charging volumes.
These findings underscore the importance of jointly optimising operational decisions of both actors, and point toward a natural extension: incorporating joint investment planning to capture long-term synergies in infrastructure sizing and tariff design. As the magnitude of the identified benefits remains highly sensitive to the local PV generation, load profiles, and mobility demand patterns, future work should assess as well the sensitivities of these synergies across a larger range of EC configurations.

\bibliographystyle{IEEEtran}
\bibliography{references}

@article{ref1a,
  title   = {A community microgrid architecture with an internal local market},
  journal = {Applied Energy},
  volume  = {242},
  author  = {B. Cornélusse and I. Savelli and S. Paoletti and A. Giannitrapani and A. Vicino},
  pages   = {547-560},
  year    = {2019},
  publisher={Elsevier}
}

@misc{REINVENT,
    title = {\uppercase{REINVENT} \uppercase{P}roject - \uppercase{E}nergy \uppercase{T}ransition \uppercase{F}und 2023 \uppercase{FPS} \uppercase{E}conomy},
    url = "https://reinvent-project.be/en" 
}

@article{velkovski2024framework,
  title={A framework for shared EV charging in residential renewable energy communities},
  author={Velkovski, Bodan and Gjorgievski, Vladimir and Markovski, Blagoja and Cundeva, Snezana and Markovska, Natasa},
  journal={Renewable Energy},
  volume={231},
  pages={120897},
  year={2024},
  publisher={Elsevier}
}

@article{allard2024quantifying,
  title={Quantifying, Activating and Rewarding Flexibility in Renewable Energy Communities},
  author={Allard, Julien and Vall{\'e}e, Fran{\c{c}}ois and De Gr{\`e}ve, Zacharie and Stegen, Thomas and Glavic, Mevludin and Corn{\'e}lusse, Bertrand},
  journal={2024 IEEE PES Innovative Smart Grid Technologies Europe (ISGT EUROPE)},
  year={2024},
  organization={IEEE}
}

@article{zanvettor2022optimal,
  title={Optimal management of energy communities hosting a fleet of electric vehicles},
  author={Zanvettor, Giovanni Gino and Casini, Marco and Giannitrapani, Antonio and Paoletti, Simone and Vicino, Antonio},
  journal={Energies},
  volume={15},
  number={22},
  pages={8697},
  year={2022},
  publisher={MDPI}
}

@book{EU,
author = {European University Institute and Nouicer, A. and Kehoe, A.-M. and Nysten, J. and Fouquet, D. and Meeus, L. and Hancher, L.},
title = {The EU clean energy package – (2020 ed.)},
publisher = {European University Institute},
year = {2020},
doi = {doi/10.2870/58299}}

@techreport{IEA_EV,
    title={Global EV Outlook},
    author = {Timur Gül and Araceli Fernandez Pales and Elizabeth Connelly},
    year ={2025},
    institution={IEA},
}

@article{machado2018overview,
  title={An overview of shared mobility},
  author={Machado, Cl{\'a}udia A Soares and de Salles Hue, Nicolas Patrick Marie and Berssaneti, Fernando Tobal and Quintanilha, Jos{\'e} Alberto},
  journal={Sustainability},
  volume={10},
  number={12},
  pages={4342},
  year={2018},
  publisher={MDPI}
}

@inproceedings{da2024electric,
  title={Electric vehicle flexibility harnessing through local energy community operation optimization: Maximizing local energy utilization},
  author={Da Silva, Enielma Cunha and Sabillon, Carlos and Venkatesh, Bala and Franco, John Fredy},
  booktitle={EEEIC/I\&CPS Europe},
  pages={1--5},
  year={2024},
  organization={IEEE}
}

@article{mouli2016system,
  title={System design for a solar powered electric vehicle charging station for workplaces},
  author={Mouli, GR Chandra and Bauer, Pavol and Zeman, Miro},
  journal={Applied Energy},
  volume={168},
  pages={434--443},
  year={2016},
  publisher={Elsevier}
}

@inproceedings{deb2017review,
  title={Review of impact of electric vehicle charging station on the power grid},
  author={Deb, Sanchari and Kalita, Karuna and Mahanta, Pinakeshwar},
  booktitle={2017 International Conference on Technological Advancements in Power and Energy (TAP Energy)},
  pages={1--6},
  year={2017},
  organization={IEEE}
}

@misc{Belpex,
  author       = {Elexys},
  title        = {Spot Belpex: Electricity spot market prices},
  howpublished = {\url{https://my.elexys.be/MarketInformation/SpotBelpex.aspx}},
  note         = {Accessed: 2025-10-14}
}

@misc{VREG,
  author       = {{Vlaamse Regulator van de Elektriciteits en Gasmarkt (VREG)}},
  title        = {Energy market in figures},
  howpublished = {\url{https://www.vlaamsenutsregulator.be/nl/energiemarkt-cijfers}},
  note         = {Accessed: 2025-10-14}
}

@article{Resflex,
	AUTHOR = {Stegen, Thomas and Castiaux, Joakim and Diffels, Noé and Duchesne, Maxime and Cornélusse, Bertrand},
	TITLE = {Residential flexible load profile generator to estimate the impact of tariffs and demand response on distribution grids and energy bills},
	LANGUAGE = {English},
	YEAR = {30 June 2025},
	SIZE = {5},
        JOURNAL = {CIRED 2025},
	LOCATION = {Geneva, Switzerland},
        BOOKTITLE= {}
}

@inproceedings{allard2025dyn,
  title={Impact of energy communities membership evolution on founding members’ expected benefits},
  author={Allard, Julien and Gasca, Mar{\'\i}a Victoria and Rigo-Mariani, Remy and Vall{\'e}e, Fran{\c{c}}ois and De Gr{\`e}ve, Zacharie and Debusschere, Vincent},
  booktitle={2025 IEEE Kiel PowerTech},
  pages={1--6},
  year={2025},
  organization={IEEE}
}
\end{document}